\documentclass[11pt,twoside]{article}

\usepackage[utf8]{inputenc}
\usepackage{natbib}

\usepackage{amsmath}
\usepackage{amssymb}
\usepackage{amsfonts}
\usepackage{latexsym}
\usepackage{textcomp}
\usepackage{pfnote}
\usepackage{fnpos}
\usepackage{dblfnote}
\usepackage{color}

\def\Xint#1{\mathchoice
{\XXint\displaystyle\textstyle{#1}}%
{\XXint\textstyle\scriptstyle{#1}}%
{\XXint\scriptstyle\scriptscriptstyle{#1}}%
{\XXint\scriptscriptstyle\scriptscriptstyle{#1}}%
\!\int}
\def\XXint#1#2#3{{\setbox0=\hbox{$#1{#2#3}{\int}$}
\vcenter{\hbox{$#2#3$}}\kern-.5\wd0}}

\def\dashint{\Xint-}

\usepackage[pdftex]{graphicx}
\usepackage{subfigure}
\usepackage{wrapfig}
\usepackage{array}
\usepackage{multicol}
\usepackage{color}
\usepackage{soul}
\usepackage{sidecap}

\usepackage{fancyhdr}
\usepackage{verbatim} 
\usepackage[pdftitle={}, pdfsubject={}, pdfkeywords={}, pdfauthor={Andr\'es F. Gutierrez.}, 
pdfproducer={Andr\'es F. Gutierrez.}, 
pdfcreator={Andr\'es F. Gutierrez-Ruiz.}, 
colorlinks=true, citecolor=blue, linkcolor=blue, anchorcolor=magenta, urlcolor=blue, pagebackref=false]{hyperref}

\title{\textbf{\huge{Innermost Stable Circular Orbits and Epicyclic Frequencies Around a Magnetized Neutron Star}}}
\author{
Andr\'es F. Guti\'errez-Ruiz$^{\text a}$, 
  C\'esar A. Valenzuela-Toledo$^{\text b}$, 
Leonardo A. Pach\'on$^{\text a}$  \\ \\   
  \small{$^{\text a}$\textit{Grupo de F\'isica At\'omica y Molecular, Instituto de F\'{\i}sica,  
  Facultad de Ciencias Exactas y Naturales,
  }} \\
  \small{\textit{Universidad de Antioquia UdeA; Calle 70 No. 52-21, Medell\'in, Colombia.}}
  \\
  \small{$^{\text b}$\textit{Departamento de F\'isica, 
  Universidad del Valle, A.A. 25360, Santiago de Cali, Colombia}}\\
  }

\date{}

\begin{document}

\maketitle

      
 \begin{abstract}
     A full-relativistic approach is used to compute the radius of the innermost stable circular 
     orbit (ISCO), the Keplerian, frame-dragging, precession and oscillation frequencies of the radial 
     and vertical motions of neutral test particles orbiting the equatorial plane of a magnetized 
     neutron star.
     The space-time around the star is modelled by the six parametric solution derived by 
    Pach\'on {\emph et al.}
     It is shown that the inclusion of an intense magnetic field, such as the one of a neutron star, have 
     non-negligible effects on the above physical quantities, and therefore, its inclusion is necessary in 
     order to obtain a more accurate and realistic description of the physical processes occurring in the 
     neighbourhood of this kind of objects such as the dynamics of accretion disk.
     The results discussed here also suggest that the consideration of strong magnetic fields may introduce
     non-negligible corrections in, e.g., the relativistic precession model and therefore on the predictions 
     made on the mass of neutron stars.
  \end{abstract}    
\section{Introducction}
\label{sec:int}
Stellar models are generally based on the Newtonian universal law of gravitation. 
However, considering the size, mass or density, there are five classes of stellar configurations 
where one can recognize significant deviations from the Newtonian theory, namely, white dwarfs, 
neutron stars, black holes, supermassive stars and relativistic star clusters \citep{MTW73}. 
In the case of magnetized objects, such as white dwarfs ($\sim10^5~$T) or neutron stars 
($\sim10^{10}~$T), 
the Newtonian theory not only fails in describing the gravitational field generated by the matter 
distribution, but also in accounting for the corrections from the energy stored in the
electromagnetic fields.

Despite this fact and due to the sheer complexity of an in-all-detail calculation of the gravitational 
and electromagnetic fields induced by these astrophysical objects, one usually appeals
to approaches based on post-Newtonian corrections \citep{AO02,Pre04}, which may or may not be 
enough in order to provide a complete and accurate description of the space-time around
magnetized astrophysical objects.
In particular, these approaches consider, e.g., that the electromagnetic field is weak compared 
to the gravitational one and therefore, the former does not affect the space-time geometry 
\citep{AO02,Pre04,Mirza04,Bakala10,Bakala12}. 
That is, it is assumed that the electromagnetic field does not contribute to the space-time curvature, 
but that the curvature itself may affect the electromagnetic field. 
%
%
Based on this approximation, the space-time around a stellar source is obtained from a
simple solution of the Einstein's field equations, such as Schwarzschild's or Kerr's solution, 
superimposed with a dipolar magnetic field \citep{AO02,Pre04,Mirza04,Bakala10,Bakala12}. 
Although, to some extend this model may be reliable for weakly magnetized astrophysical sources, 
it is well known that in presence of strong magnetic fields, non-negligible contributions to the 
space-time curvature are expected, and consequently on the physical parameters that describe 
the physics in the neighbourhood of these objects.

In particular, one expects contributions to the radius of the innermost stable circular orbit (ISCO)
\citep{Sanabria10}, the Keplerian frequency, frame-dragging frequency, precession and oscillation 
frequencies of the radial and vertical motions of test particles \citep{Bakala10,Bakala12,Stella99}, 
and perhaps to other physical properties such as the angular momentum of the emitted radiation 
\citep{Tamburini11}, which could reveal some properties of accretion disks and therefore of the compact 
object \citep{Bocquet95,Broderick00,Konno99,Cardall00}. 
In other words, to construct a more realistic theoretical description that includes purely relativistic effects
such as the modification of the gravitational interaction by electromagnet fields, it is necessary to use a 
complete solution of the full Einstein-Maxwell field equations that takes into account all the possible 
characteristics of the compact object.
In this paper, we use the six parametric solution derived by \cite{PRS06} (hereafter
PRS solution), which provides an adequate and accurate description of the exterior field of a 
rotating magnetized neutron star \citep{PRS06,PRV12}, to calculate the radius of the ISCO, the Keplerian, 
frame-dragging, precession and oscillation frequencies of neutral test-particles orbiting the equatorial 
plane of the star.
The main purpose of this paper is to show the influence of the magnetic field on these particular 
quantities.

The paper is organized as follows. In section \ref{sec:des}, we briefly describe the physical properties 
of the PRS solution, the general formulae to calculate the parameters  that characterize 
the dynamics around the star are presented in section \ref{sec:des2}. 
Sections \ref{sec:isco} and \ref{sec:klep} are devoted to the study of the influence of the magnetic field 
on the ISCO radius and on the Keplerian and epicyclic frequencies, respectively. 
The effect of the magnetic field on the energy $E$ and the angular momentum $L$ is outlined in section \ref{sec:others}.
Finally, the conclusions of this paper are given in section \ref{sec:con}.
    

\section{Description of the Space-Time Around the Source}
\label{sec:des}
According to  \cite{1953AnP...447..309P}, the metric element $ds^2$ around a rotating object with 
stationary and axially symmetric fields can be cast as
\begin{align}
 ds^2 &=-f(dt-\omega  d\phi)^2\nonumber
+
 f^{-1}[e^{2 \gamma}(d\rho^2+dz^2)+\rho^2d\phi^2],\label{line}
\end{align}

where $f$, $\gamma$ and $\omega$ are functions of the quasi-clyndrical Weyl-Papapetrou coordinates 
$(\rho, z)$. The non-zero components of metric tensor, which are related to the metric functions $f$, $
\omega$ and $\gamma$, are
\begin{align}
g_{\phi \phi} &= \frac{\rho^2}{f(\rho,z)} - f(\rho,z) \omega(\rho,z)^2,
\qquad
g_{tt} = -f(\rho,z),
\\
g_{t\phi} &= f(\rho,z) \omega(\rho,z),
\qquad
g_{zz} = g_{\rho \rho} = \frac{{\rm e}^{2\gamma(\rho,z)}}{f(\rho,z)}
= \frac{1}{g^{zz}} =  \frac{1}{g^{\rho\rho}}.
\end{align}

By using the Ernst procedure and the line element in equation (\ref{line}), it is possible to rewrite the Einstein-
Maxwell equations  in terms of two complex potentials ${\cal{E}}(\rho, z)$ and $\Phi(\rho,z)$ [see 
\cite{1968PhRv..172.1850E} for details]
\begin{eqnarray}
({\rm Re}\,{\cal E}+|\Phi|^2)\nabla^2{\cal E}&=& (\nabla {\cal E} +
2\Phi^*\nabla\Phi)\cdot\nabla{\cal E}, 
\nonumber\\ 
({\rm Re}\,{\cal
E}+|\Phi|^2)\nabla^2\Phi&=& (\nabla {\cal E} +
2\Phi^*\nabla\Phi)\cdot\nabla\Phi\, ,\label{Ernst}
\end{eqnarray}
where $^*$ stands for complex conjugation.
The above system of equations can be solved by means of the Sibgatullin integral method 
\citep{1991owsg.book.....S,1993CQGra..10.1383M}, according to which the Ernst potentials 
can be expressed as
\begin{eqnarray}
\label{ints}
\nonumber
    {\cal E}(z,\rho)=\frac{1}{\pi}\int_{-1}^1
    \frac{e(\xi)\mu(\sigma)d\sigma}{\sqrt{1-\sigma^2}},
\qquad
    \Phi(z,\rho)=\frac{1}{\pi}\int_{-1}^1
    \frac{f(\xi)\mu(\sigma)d\sigma}{\sqrt{1-\sigma^2}}\, ,
\end{eqnarray}
where $\xi = z + \mathrm{i} \rho \sigma$ and $e(z)={\cal E}(z,\rho=0)$ and $f(z)=\Phi(z,\rho=0)$
are the Ernst potentials on the symmetry axis.
As shown below, these potentials contain all information about the multipolar structure 
of the astrophysical source [see also \cite{2006CQGra..23.3251P} for a discussion on the
symmetries of these potentials]. 
The auxiliary unknown function $\mu(\sigma)$ must satisfy the integral and normalization 
conditions
\begin{equation}
\dashint_{-1}^{1}\frac{\mu(\sigma)[e(\xi)+\tilde
e(\eta)+2f(\xi)\tilde
f(\eta)]d\sigma}{(\sigma-\tau)\sqrt{1-\sigma^2}}=0,
\qquad
\int_{-1}^1\frac{\mu(\sigma)d\sigma}{\sqrt{1-\sigma^2}}=\pi.
\end{equation}
Here $\eta=z+i\rho\tau$ and $\sigma, \tau\in[-1,1]$, 
$\tilde e(\eta)=e^*(\eta^*)$, $\tilde f(\eta)=f^*(\eta^*)$.

For the PRS solution \citep{PRS06}, the Ernst potentials were chosen as:
\begin{eqnarray}
\label{axispot}
e(z) = \frac{z^3-z^2(m+ia)-kz+is}{z^3+z^2(m-ia)-kz+is}\, ,
\qquad
\label{Ernstaxis} 
f(z) = \frac{qz^{2} + i\mu z}{z^3+z^2(m-ia)-kz+is}.
\end{eqnarray}
The electromagnetic and gravitational multipole moments of the source were calculate
by using the Hoenselaers \& Perj\'es method \citep{Hoenselaers90} and are given by
\citep{PRS06}:
\begin{align}
\label{multipolarmass}
     M_0 &= m ,\quad M_2 =(k-a^2)m,...
\qquad \hspace{0.39cm}
    S_{1}=a  ,\quad S_3=- m(a^3 - 2 a k + s),...
\\
\label{multipolarscurrent}
     Q_{0} &= q, \quad Q_{2}=-a^{2}q-a\mu+kq,...
\quad
  B_{1} = \mu+aq,...
\quad
 B_{3} =-a \mu +\mu k-a^{3}q +2akq-qs,...
\end{align}
where the $M_i$s denote the moments related to the mass distribution and $S_i$s to the
current induced by the rotation.
Besides, the $Q_i$s are the multipoles related to the electric charge distribution and the 
$B_i$s to the magnetic properties. 
In the previous expressions,  $m$ corresponds to the total mass, $a$ to the total angular moment 
per unit mass ($a$ = $J/M_0$, being $J$ the total angular moment), $q$ to the total electric 
charge.
In our analysis, we set the electric charge parameter $q$ to zero because, as it is case of 
neutron stars, most of the astrophysical objects are electrically neutral.
The parameters $k, \ s$ and $\mu$ are related to the mass quadrupole moment, the current 
octupole, and the magnetic dipole, respectively.



Using Eqs.~(\ref{ints}) and (\ref{axispot}), the Ernst potentials obtained by \cite{PRS06} are
\begin{equation}\label{potenciales_ernst}
{\cal E}=\frac{A + B }{A - B}\, , \qquad \Phi=\frac{C}{A - B}\, ,
\end{equation}
which leads to the following metric functions:
\begin{align}
\label{eq:metricfuncs}
f&=\frac{A \bar{A}-B \bar{B} + C \bar{C}}{( A -
B)(\bar{A}-\bar{B})}\, ,\quad e^{2\gamma}=\frac{A \bar{A} -B
\bar{B} + C \bar{C}}{\displaystyle{K \bar{K}\prod_{n=1}^{6}r_n}}\, ,\\
\omega &= \frac{{\rm Im}[(A + B)\bar{H}-(\bar{A} + \bar{B})G - C
\bar{I}]}{A \bar{A} - B \bar{B} + C \bar{C}}.
\end{align}
The analytic expressions for the functions $A$, $B$, $C$, $H$, $G$, $K$, and $I$ can be found 
in the original reference \citep{PRS06} or in Appendix of \cite{PRV12}.
A Mathematica 8.0 script with the numerical implementation of the solution can be found at
\href{http://gfam.udea.edu.co/~lpachon/scripts/nstars}
{http://gfam.udea.edu.co/$\sim$lpachon/scripts/nstars}.
\section{Characterization of the Dynamics Around the Source}
\label{sec:des2} 
In the framework of general relativity, the dynamics of a particle may be analyzed via the Lagrangian 
formalism as follows. Let us consider a particle of rest mass $\mathfrak{m}_0=1$ moving in a space-time 
characterized by the metric tensor $g_{\mu\nu}$, thus the Lagrangian of motion of the particle is 
given by
\begin{equation}
\label{lag}
{\cal L}=\frac{1}{2}g_{\mu\nu}\dot{x}^\mu\dot{x}^\nu,
\end{equation}
where the dot denotes differentiation with respect to the proper time $\tau$, $x^\mu(\tau)$ are 
the coordinates. Since the fields are stationary and axisymmetric, there are two constants of motion 
related to the time coordinate $t$  and azimuthal coordinate $\phi$ [see \cite{1995PhRvD..52.5707R} 
for details], these are given by:
%
%
\begin{equation}
\label{ene}
 E=-\frac{\partial {\cal L}}{\partial \dot{t}}=
 -g_{tt}\bigg(\frac{dt}{d\tau}\bigg)-g_{t\phi}\bigg(\frac{d\phi}{d\tau}\bigg),
\end{equation}
\begin{equation}
\label{ang}
L=\frac{\partial {\cal L}}{\partial \dot{\phi}}=
g_{t\phi}\bigg(\frac{dt}{d\tau}\bigg)+g_{\phi\phi}\bigg(\frac{d\phi}{d\tau}\bigg),  
\end{equation}
where $E$ and $L$ are the energy and the canonical angular momentum per unit mass, respectively. 
For real massive particle, the four velocity is a time-like vector with normalization $g_{\mu\nu}u^\mu u^\nu=-1$.
If the motion takes place in the equatorial plane of the source $z=0$, this normalization condition leads to
\begin{align}
\label{norm}
g_{\rho \rho}\bigg(\frac{d\rho}{d\tau}\bigg)^{2}= -1 + g_{tt} \bigg(\frac{dt}{d\tau}\bigg)^{2} 
+ g_{\phi \phi} \bigg(\frac{d\phi}{d\tau}\bigg)^{2} 
+ 2g_{t\phi}\bigg(\frac{dt}{d\tau}\bigg) \bigg(\frac{d\phi}{d\tau}\bigg).
\end{align}

From equation (\ref{norm}), one can identify an effective potential that governs the geodesic 
motion in the equatorial plane [see e.g. \cite{1972ApJ...178..347B}]
\begin{equation}
\label{(9)}
 V_{\mathrm{eff}}(\rho)=1-\frac{E^2g_{\phi \phi}+2ELg_{t\phi}
 +L^2 g_{tt}}{g^2_{t\phi}-g_{\phi \phi}g_{tt}}.
\end{equation}
For circular orbits, the energy and the angular momentum per unit mass, are determined 
by the conditions $V_{\mathrm{\mathrm{eff}}}(\rho)=0$ and $dV_{\mathrm{eff}}/d\rho=0 $.
Based on these conditions, one can obtain expressions for the energy and the angular 
momentum of a particle that moves in a circular orbit around the star [see e.g. 
\cite{2002MNRAS.336..831S}], namely,
\begin{equation}
\label{(4)}
 E=\frac{\sqrt{f}}{\sqrt{1-f^2\chi^2/\rho^2}},
\qquad
L=E(\omega + \chi),
 \end{equation}
 \begin{equation}
\label{(5)}
 \chi=\left\{\rho[-\omega_{,\rho}f^2-\sqrt{\omega^2_{,\rho}f^4+f_{,\rho}\rho(2f-f_{,\rho}\rho)}\right\}/
 \left[f(2f-f_{,\rho}\rho)\right],
 \end{equation}
where the colon stands for a partial derivative respect the lower index. 
The radius of innermost stable circular orbit (ISCO's radius) is determined by solving 
numerically for $\rho$ the equation 
 $$d^2V_{\mathrm{eff}}/d\rho^2=0,$$
which arising from the marginal stability condition.
This condition together with the equations (\ref{(4)}) and (\ref{(5)}) can be write in terms of the metric functions 
as [see \cite{2002MNRAS.336..831S}]
{\small
\begin{eqnarray}\label{isco}
\omega_{,\rho}\omega_{,\rho \rho}f^5\rho(2f-f_{,\rho}\rho)+\omega^2_{,\rho}f^4\big[2f^2+(f_{,\rho \rho}f-f_{,\rho})\rho^{2}\big]
+ \omega_{\rho}f^2\sqrt{\omega^2_{,\rho}f^4+f_{,\rho}(2f-f_{,\rho}\rho)}\big[2f^2-f\rho(4f_{,\rho}+f_{,\rho \rho} \rho)
&&\nonumber\\
+2f^2_{,\rho}\big]+\rho(2f-f_{,\rho}\rho)\Big[3f_{\rho}f^2-4f^2_{, \rho}f\rho+f^3_{,\rho}\rho^2 
+f^2\big[f_{\rho \rho}\rho-\omega_{,\rho \rho} f \sqrt{\omega^2_{,\rho}f^4+f_{,\rho}(2f-f_{,\rho}\rho)}\big]\Big]\ = \ 0.
\end{eqnarray}
}
As is usual  in the literature, the physical ISCO radius reported here corresponds to evaluation 
of $\sqrt{g_{\phi \phi}}$ at the root of equation (\ref{isco}). 
This equation is solved for fixed total mass of the star $M$, the dimensionless spin parameter $j=J/M^2$ 
(being $J$ the angular momentum), the quadrupole moment $M_2$ and the magnetic dipolar moment
$\mu$ [see Table VI of \cite{Pappas12a}].

\subsection{Keplerian, Oscillation and Precession Frequencies}
The Keplerian frequency $\Omega_{\mathrm K}$ at the ISCO can be obtained from the equation 
of motion of the radial coordinate $\rho$. This equation is easily obtained by using the Lagrangian 
(\ref{lag}),
\begin{align}
g_{\rho\rho}\ddot{\rho} -\frac{1}{2}\Big[- g_{\rho\rho,\rho} 
\dot{\rho}^2 +g_{\phi\phi,\rho} \dot{\phi}^2 + 
g_{tt,\rho} \dot{t}^2+g_{t\phi,\rho} \dot{t}\dot{\phi} \Big]&=0.
\end{align}
By imposing the conditions of circular orbit or constant orbital radius, $d\rho/d\tau=0$ and $d^2\rho/d
\tau^2=0$ and taking into account that
$d\phi/d\tau=\Omega_K dt/d\tau$, one gets [see e.g. \cite{1995PhRvD..52.5707R}]
\begin{equation}
\label{8}
 \Omega_{\mathrm{K}}=\frac{d\phi}{dt}=
 \frac{-g_{t \phi,\rho} \pm \sqrt{(g_{t\phi,\rho})^2-g_{\phi \phi,\rho}g_{tt,\rho}}}{g_{\phi \phi,\rho}},  
\end{equation}
where ``$+$'' and ``$-$''  denotes the Keplerian frequency for corotating and counter-rotating 
orbits, respectively.

The epicyclic frequencies are related to the oscillations frequencies of the periastron and 
orbital plane of a circular orbit when we apply a  slightly radial and vertical perturbations to it. 
According to \cite{1995PhRvD..52.5707R}, the radial and vertical epicyclic frequencies are 
given by the expression
\begin{align}
\label{9}
 \nu_{\alpha}&= \frac{1}{2\pi}\bigg\{-\frac{g^{\alpha\alpha}}{2}\bigg[(g_{tt}
 +g_{t\phi}\Omega_{k})^2\bigg(\frac{g_{\phi\phi}}{\rho^2}\bigg)_{,\alpha\alpha} \nonumber
 \\
 &-
 2(g_{tt}+g_{t\phi}\Omega_{k})(g_{t\phi}+g_{\phi\phi}\Omega_{k})\bigg(\frac{g_{t \phi}}{\rho^2}\bigg)_{,\alpha\alpha} \nonumber
+
(g_{t\phi}+g_{\phi\phi}\Omega_{k})^2\bigg(\frac{g_{tt}}{\rho^2}\bigg)_{,\alpha\alpha}\bigg]\bigg \},
\end{align}
where $\alpha=\{\rho,z\}$.
The periastron $\nu_\rho^{\mathrm p}$ and the nodal $\nu_z^{\mathrm p}$ frequencies are defined by:
\begin{equation}\label{10}
 \nu^{\mathrm p}_{\alpha}=\frac{\Omega_{k}}{2\pi}-\nu_{\alpha},
\end{equation}
which are the ones that have an observational interest \citep{Stella99}. 
The radial oscillation frequency vanishes at the ISCO radius and therefore, the radial precession 
frequency equals to the Keplerian frequency.

Finally, the frame dragging precession  frequency or Lense--Thirring frequency  
$\nu_{\mathrm{LT}}$ is given 
by [see e.g. \cite{1995PhRvD..52.5707R}]
\begin{equation}
\nu_{\mathrm{LT}}=-\frac{1}{2\pi}\frac{g_{t\phi}}{g_{\phi\phi}}\ .
\end{equation}
$\nu_{\mathrm{LT}}$ is related to purely relativistic effects only \citep{MTW73}.
In this phenomenon, the astrophysical source drags the test particle into the direction of 
its rotation angular velocity.
In absence of electromagnetic contributions, as in the case of the Kerr solution, the frame dragging
comes from the non-vanishing angular momentum of the source.
In the presence of electromagnetic contributions in non-rotating sources, it was shown by 
\cite{herrera2006} that the non-zero circulation of the Poynting vector is able to induced frame 
dragging, in which case it is \emph{electromagnetically induced}.
In the cases discussed below, due to the presence of fast rotations and magnetic fields, the frame 
dragging will be induced by a combination of these two processes.

\section{Influence of the Dipolar Magnetic Field in the ISCO Radius.}
  \label{sec:isco} 
As it was discussed in the Introduction, contributions from the energy stored in the electromagnetic 
fields come via equivalence between matter and energy, $E = m c^2$.
For the particular case of a magnetar ($\sim10^{10}~$T), the electromagnetic energy density is around 
$4\times10^{25}~$J/m$^3$, with an $E /c^2$ mass density $10^4$ times larger than that of lead.
Hence, the relevant physical quantities should depend upon the magnetic field, and in particular on the 
magnetic dipole moment $\mu$.

\begin{table*}
\begin{center}
\begin{tabular}{ccccc}\hline\hline
 Model       & $M_0$ [km] & $j$  & $M_{2}$ [km$^3$]& $S_3$ [km$^4$] \\ \hline\hline
M17 &  4.120 & 0.588 &  -51.8  &-210.0 \\ 
M18 &  4.139 & 0.635 & -62.6   &-279.0 \\ 
M19 &  4.160 & 0.682 & -74.9   & -365.0 \\ 
M20 &  4.167 & 0.700 &  -79.8  & -401.0 \\\hline\hline
\end{tabular}
\end{center}
\caption{Realistic numerical solutions for rotating neutron stars derived by  \cite{Pappas12a}. 
Here, $M_0$ is the total mass of the star, $j$ is the dimensionless spin parameter: $j=J/M_0^2$ 
(being $J$ the angular momentum), $M_2$ is the quadrupole moment and $S_3$ is the current 
octupole moment [see Table VI of \cite{Pappas12a}].}
\label{tp}
\end{table*}
\begin{figure}[!h]
	   \centering \includegraphics[ width=0.5\columnwidth]{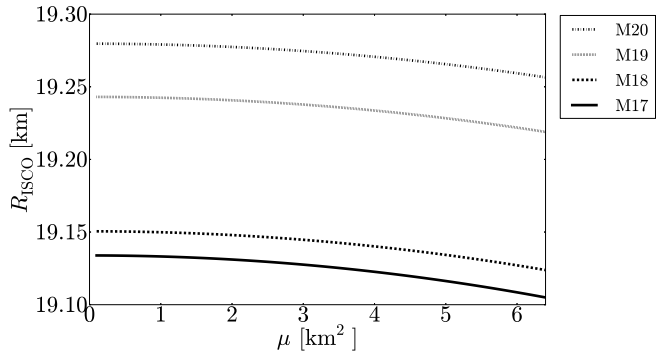} 
  \caption{
               \small ISCO radius as a function of magnetic dipole parameter $\mu$. 
	     The physical parameters for the star correspond to the models M17-M20 listed in Table \ref{tp}. 
	     An increase of $\mu$ leads to a decrease of the ISCO radius.
	      }
\label{ROME}	      
\end{figure}

Figure~\ref{ROME} shows the ISCO radius as a function of the parameter $\mu$ for four 
particular realistic numerical solutions for rotating neutron stars models derived by \cite{Pappas12a}. 
The models used coincide with the models 17,18,19 and 20 of Table VI in that reference and
correspond to results for the Equation of State L (see Table~\ref{tp}). 
The lowest multipole moments of the PRS solution, namely, mass, angular moment and mass 
quadrupole were fixed to the numerical ones obtained by \cite{Pappas12a} (see Table~\ref{tp}). 
Since the main objective here is to analyze the influence of the magnetic dipole, the current octupole 
parameter $s$ was set to zero. 
Note that this does not mean that the current octupole moment vanishes (see in Eq.~(\ref{multipolarmass}). 
The parameter $\mu$ was varied between $0$ and $6.4$~km$^2$, which corresponds
to magnetic fields dipoles around $0$ and $6.3\times10^{31}$~Am$^2$, respectively. 
In all these cases, the ISCO radius decreases for increasing $\mu$, this can be understood as
a result of the dragging of inertial frames induced by the presence of the magnetic dipole,
we elaborate more on this below.

\section{Keplerian and Epicyclic Frequencies}
\label{sec:klep} 
Some of the predictions from General Relativity (RG) such as the dragging of inertial 
systems (frame-dragging or Lense-Thirring effect) \citep{2011PhRvL.106v1101E}, the geodesic 
precession (geodesic effect or de Sitter precession)  \citep{2011PhRvL.106v1101E} or the analysis 
of the periastron precession of the orbits \citep{2010PhRvL.105w1103L} have been experimental 
verified.
However, since they have observed in the vicinity of the Earth, these observations represent 
experimental support to RG only in the weak field limit.
Thus, it is fair to say that the RG has not been checked in strong field limit [see e.g. 
\cite{2008LRR....11....9P}]. 
In this sense, the study of compact objects such as black holes, neutron stars, magnetars, etc.,
is certainly a topic of great interest, mainly, because these objects could be used as remote laboratories 
to test fundamental physics, in particular, the validity of general relativity in the strong field limit
yet \citep{Stella99,PRV12}.
\\ \\
The relevance of studying the Keplerian, epicyclic and Lense-Thirring frequencies relies on the fact that 
they are usually used to explain the quasiperiodic oscilation phenomena present in some Low Mass 
X-ray Binaries (LMXRBs). In this kind of systems a compact object accretes matter form the other one 
(which is usually a normal star) and the X-ray emissions could be related to the relativistic motion of 
the accreted matter e.g. rotation, oscillation and precession. 
\cite{Stella99} showed that in the slow rotation regime, the periastron precession
and the Keplerian frequencies could be related to the phenomena of the kHz quasiperiodic 
oscillations (QPOs) observed in many accreting neutron stars in LMXRBs \citep{Stella99}. 
This model is known as the Relativistic Precession Model (RPM) \citep{Stella99} and identifies the 
lower and higher QPOs frequencies with the periastron precession and the keplerian frequencies respectively. 
The RPM model has been used to predict the values of the mass and angular momentum of the 
neutron stars in this kind of systems [see e.g. \cite{Stella99} and \cite{1999ApJ...524L..63S}, for details]. 

\begin{figure*}[!h]
   	  \centering
	  \subfigure[{\label{FR3a}}]{\includegraphics[width=0.45\columnwidth]{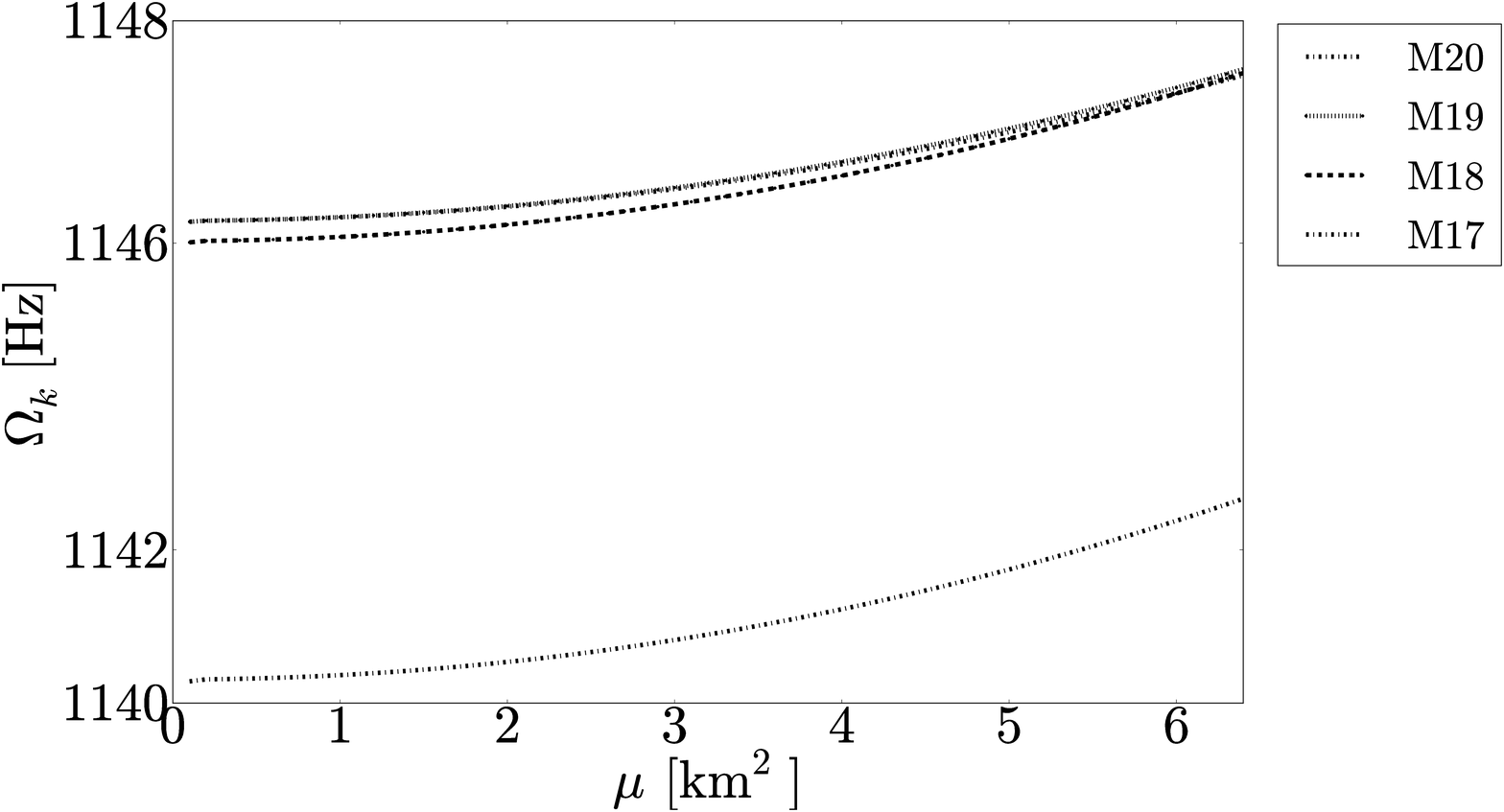}}
       \subfigure[{\label{FR3b}}]{\includegraphics[width=0.45\columnwidth]{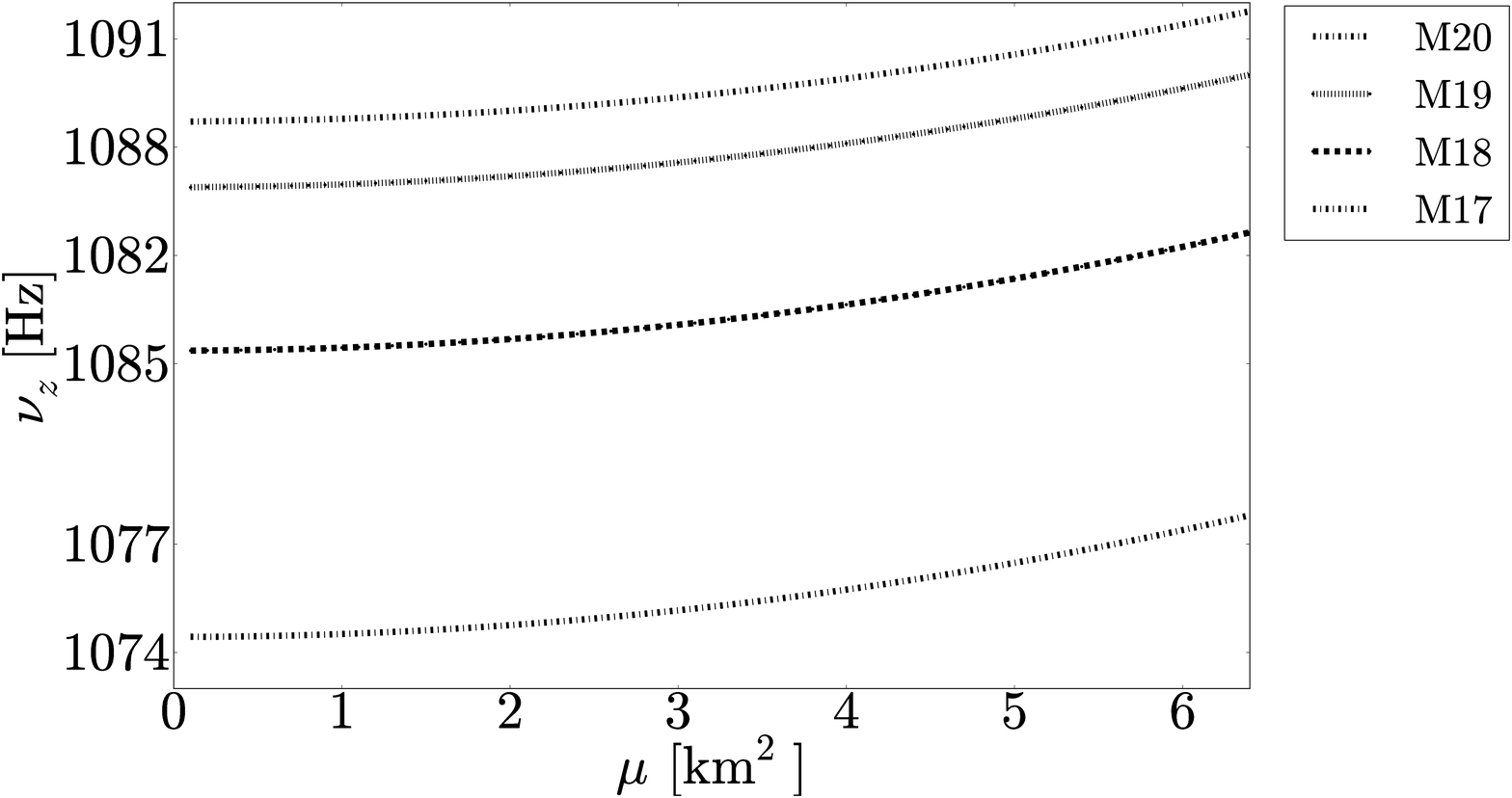}}
	     \subfigure[{\label{FR3c}}]{\includegraphics[width=0.45\columnwidth]{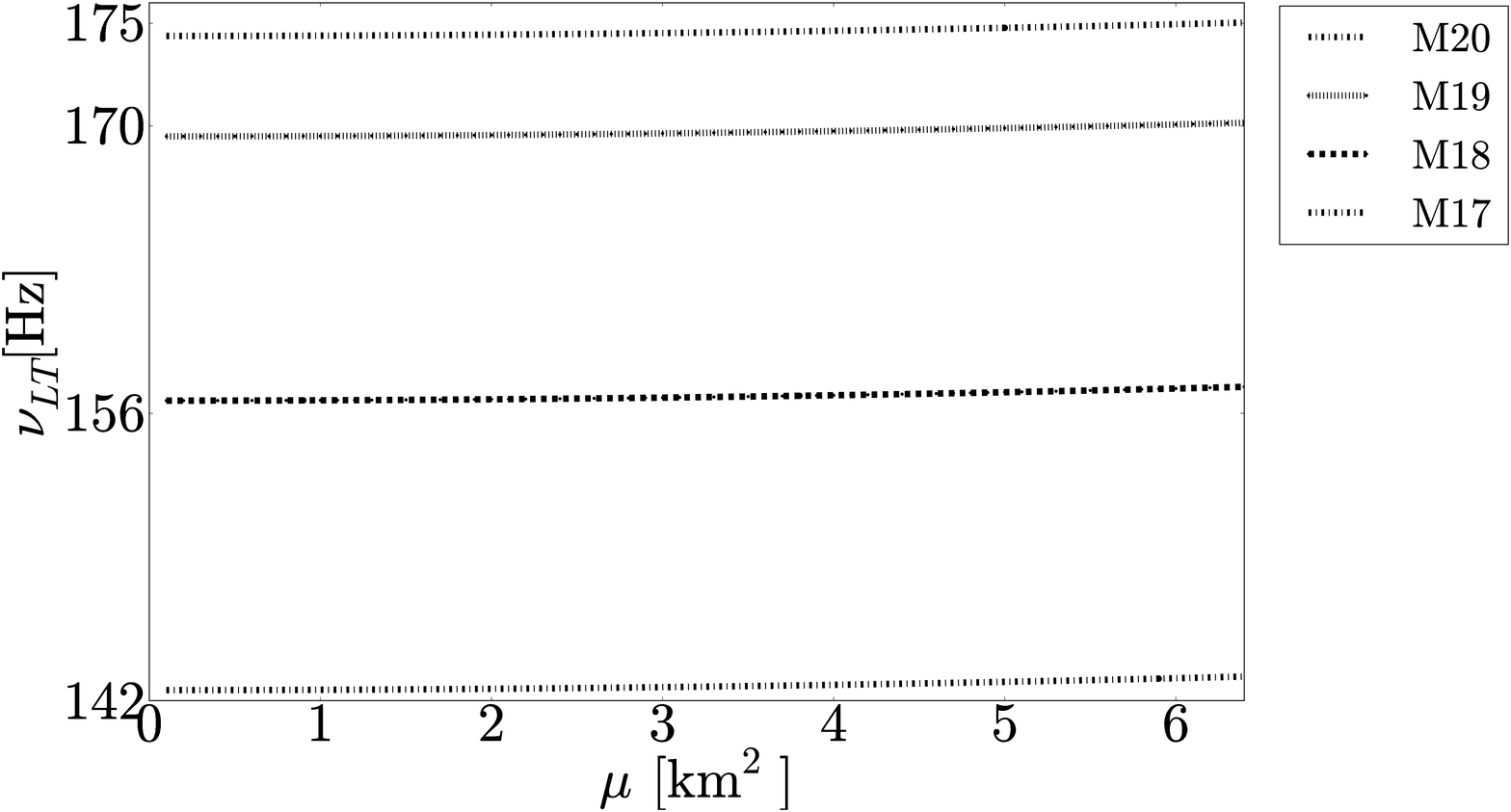}}
  \caption{
                 \small  Keperian frequency (a), nodal precession frequency (b) and, Lense-Thirring 
                 frequency (c) as a function of the $\mu$ parameter. 
                 All the frequencies increase for increasing $\mu$. }
\label{FR3}	      
 \end{figure*}

In the framework of the RPM, it is assumed that the motion of the accretion disk is determined by 
the gravitational field alone and thereby, it is normal to assume that the exterior gravitational field 
of the neutron star is well described by the Kerr metric. 
However, most of the neutron stars have (i) quadrupole deformations that significantly differ 
from Kerr's quadrupole deformation \citep{1999ApJ...512..282L} and (ii) a strong magnetic field.
The influence of the non-Kerr deformation was discussed, e.g., by \cite{2010ApJ...716..187J} and 
\cite{PRV12}. 
Based on observational data, it was found that non-Kerr deformations dramatically affect, 
e.g., predictions on the mass of the observed sources.
Below, we consider the influence of the magnetic field and find that it introduces non-negligible 
corrections in the precession frequencies and therefore, on the predictions made by the Kerr-based 
RPM model. 

In the previous section, it was shown that the magnetic field affects  the value of the ISCO radius 
of a neutral test particle that moves around the exterior of a magnetized neutron star. 
Below it shown that other physicals quantities such as the Keplerian and epicyclic frequencies, 
which are used to describe the physics of accretion disk, are also affected. 
The influence on the Lense-Thirring frequency is also discussed. 

Fig.~\ref{FR3} shows the influence of the magnetic field in the Keplerian [Fig.~\ref{FR3a}], 
nodal precession [Fig.~\ref{FR3b}] and Lense-Thirring [Fig.~\ref{FR3c}] frequencies. 
In all cases, the frequencies are plotted as a function of the parameter $\mu$ while fixing the others 
free parameters (mass, angular moment and mass quadrupole) according to the realistic numerical 
solutions in Table~\ref{tp}. 
As it expected is for shorter ISCO radius (see previous section), all frequencies increase with 
increasing $\mu$.
%
The changes in the Lense-Thirring frequency come from the electromagnetic contribution
discussed above.
%
 

\section{Energy and Angular Momentum}
\label{sec:others} 
\begin{figure*}[t]
   	    \centering
	    \subfigure[{\label{Fig. 3a}}]{\includegraphics[width=0.45\columnwidth]{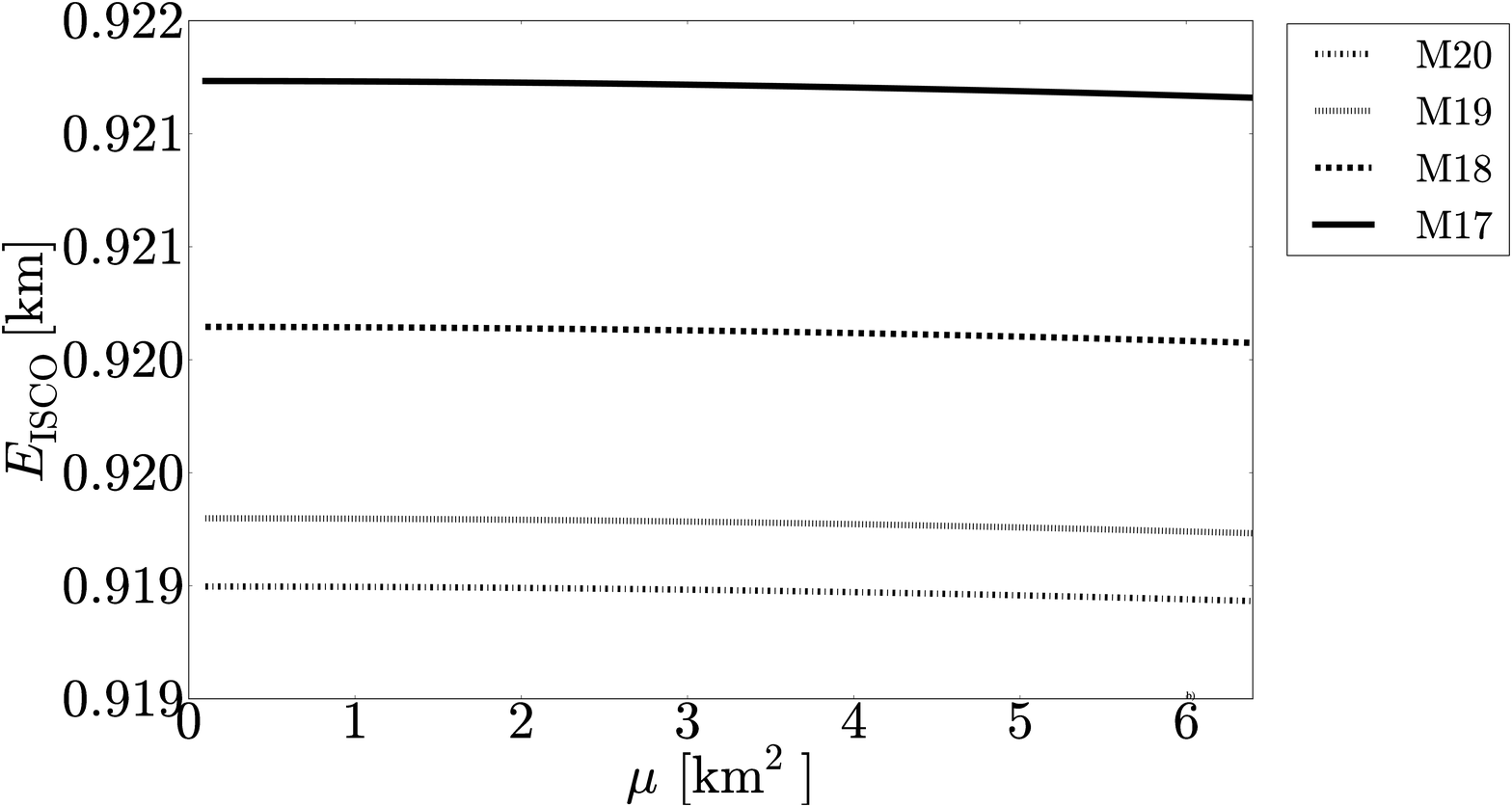}}  
	      \subfigure[{\label{Fig. 3b}}]{\includegraphics[width=0.45\columnwidth]{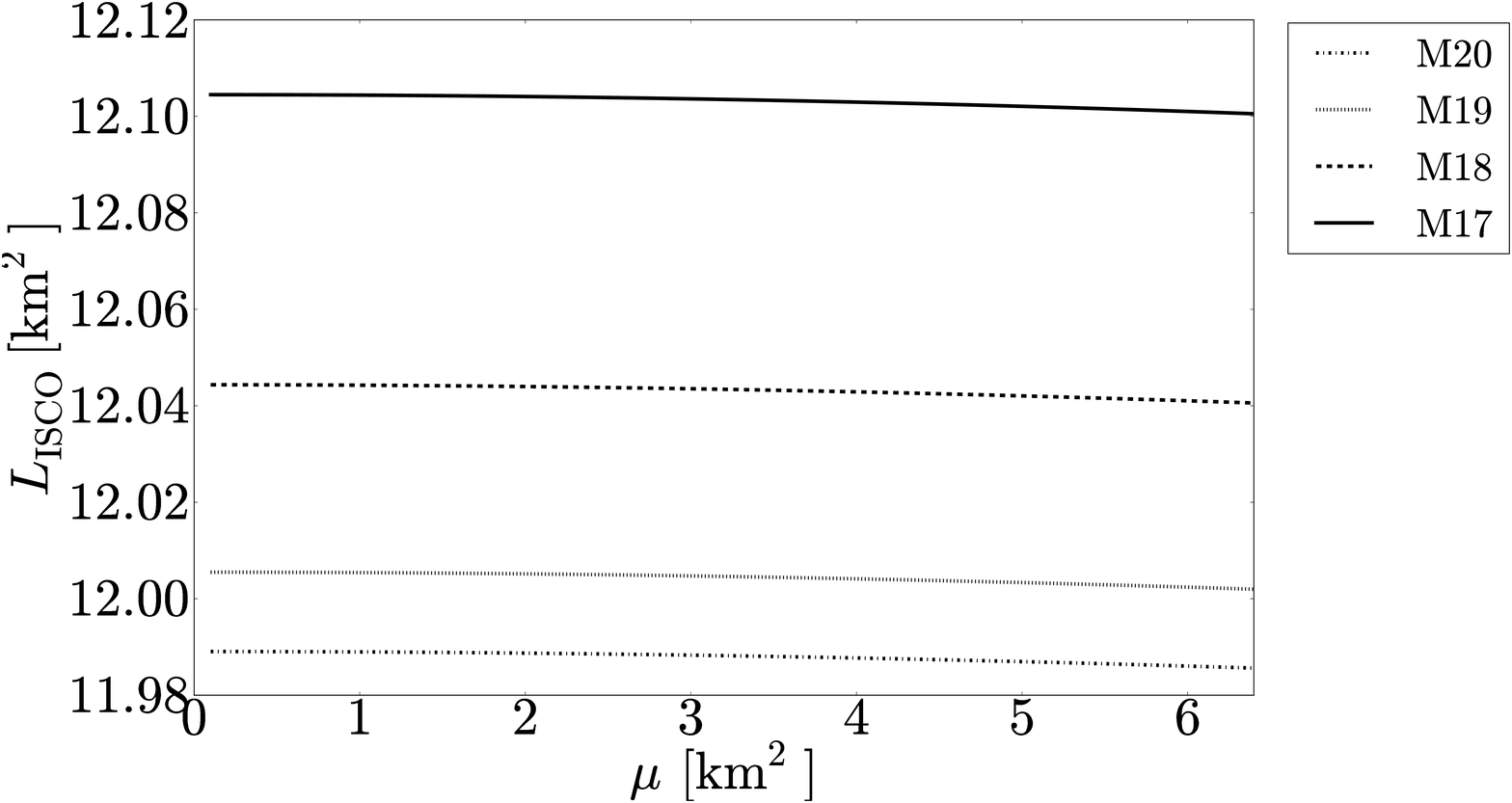}} 
		\caption{\small Energy (a) and angularmomentum (b) of a test particle at the ISCO radius 
		versus magnetic dipole parameter 
	     $\mu$ given by the PRS solution. 
	     We can see that reduction of the energy and angular momentum 
	     of the particle while the the magnetic dipole of the star is increased.}
 \end{figure*}

\begin{figure*}[t]
   	    \centering
	    \subfigure[{\label{Fig. 4a}}]{\includegraphics[width=0.45\columnwidth]{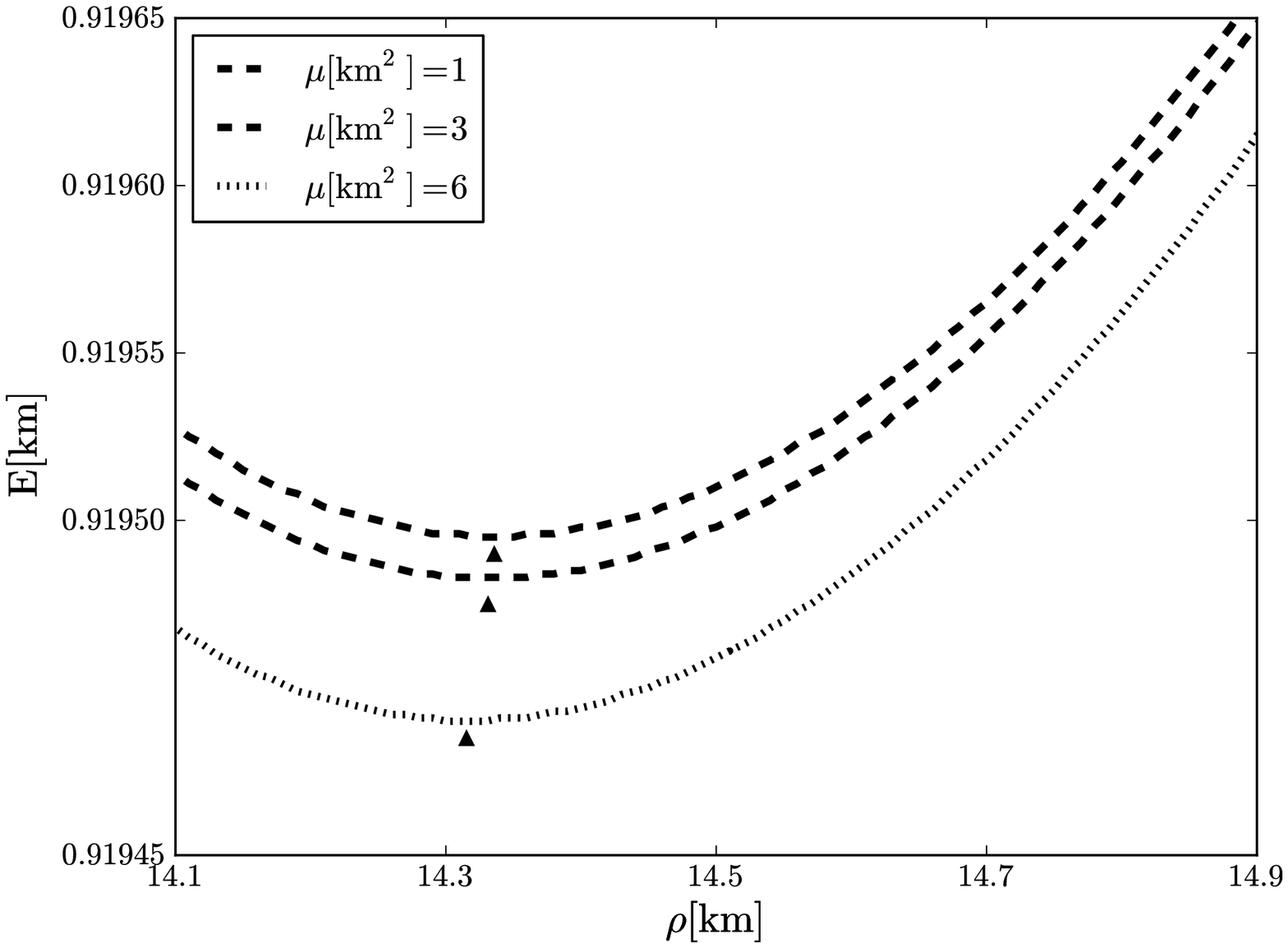}}  
	      \subfigure[{\label{Fig. 4b}}]{\includegraphics[width=0.45\columnwidth]{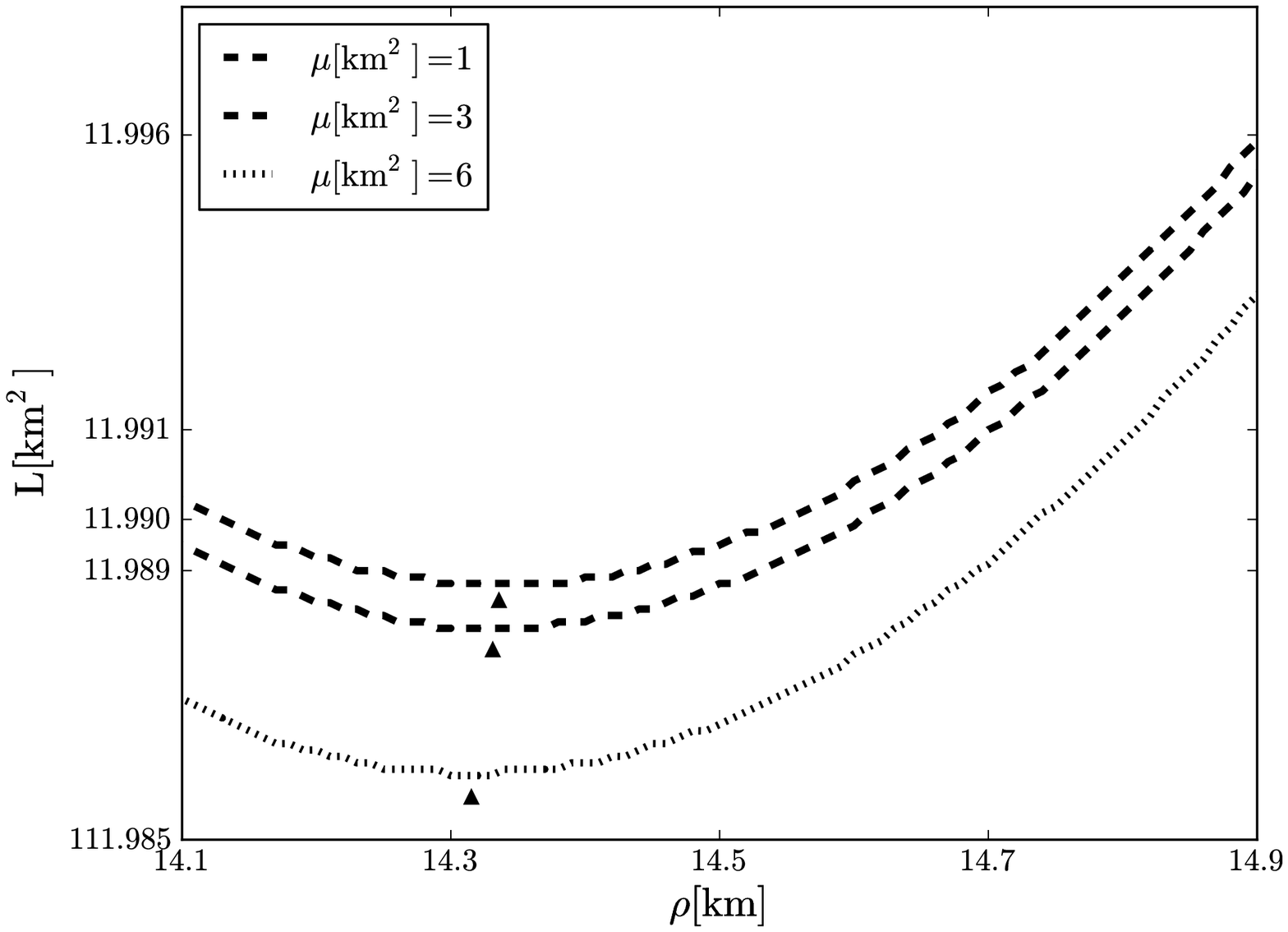}} 
		\caption{\small Energy (a) and angular momentum (b) of a test particle at the ISCO radius 
		versus magnetic dipole parameter 
	     $\mu$ given by the PRS solution. 
	     We can see that reduction of the energy and angular momentum 
	     of the particle while the the magnetic dipole of the star is increased.}
 \end{figure*}

In order to understand the results presented above, we consider that it is illustrative to consider
first the effect that the magnetic field has on the energy $E$ and the angular momentum 
$L$ needed to described marginally stable circular orbits [see equations (\ref{(4)}) and (\ref{(5)})].
In doing so, we fix the total mass $M_0$, the spin parameter $j$ and the quadrupole moment $M_2$, 
according to the values in Table~\ref{tp}. 
Figures~\ref{Fig. 3a} and \ref{Fig. 3b} show $E$ and $L$ at the ISCO as a function of the dipolar 
moment $\mu$.
By contrast to the case of the Keplerian frequency, an increase of $\mu$ decreases the value of the 
energy and the angular momentum needed to find an ISCO.
Complementarily, in figures~\ref{Fig. 4a} and \ref{Fig. 4b}, $E$ and $L$ are depicted as a function of 
$\rho$ for various values of the dipolar moment.
Since the energy and the angular momentum associated to the ISCO correspond to the minima of 
the curves $E(\rho)$ and $L(\rho)$ (indicated by triangles), the figures show that an increase of the 
magnetic dipole moment induces a decrease of the ISCO radius (see Fig.~\ref{ROME} above) and 
simultaneously a decrease of the energy and the angular momentum.

At a first sight, for an increasing magnetic field, it may seem conspicuous that the angular momentum,
of co-rotating test particles, decreases [see Fig.~\ref{Fig. 4b}] whereas the Keplerian frequency 
increases [see Fig. \ref{FR3a}]. 
However, this same opposite trend is already present in the dynamics around a Kerr source when 
the angular momentum of the source is increased [cf.~equations (2.13) for the angular momentum 
and (2.16) for the Keplerian frequency in \cite{1972ApJ...178..347B}].
In Kerr's case, the co-rotating test particles are dragged toward the source thus inducing a shorter 
ISCO radius [cf. equation~(2.21) in \cite{1972ApJ...178..347B}], and since the leading order in the 
Keplerian frequency goes as $\sim1/\rho^2$, a larger frequency is expected.
By contrast, the contra-rotating test particles are ``repelled'' by the same effect thus resulting in an 
increase of the ISCO radius and in a decrease of the Keplerian frequency [see also Fig.~2 in 
\cite{PRV12}].

Having in mind the situation in Kerr's case and by noting that the frame dragging can be induced 
by current multipoles of any order \citep{herrera2006}, the goal now is to compare the characteristics 
of the contributions from the angular momentum and the dipole moment to the Keplerian frequency 
based on the approximate expansions derived by \cite{Sanabria10} and appeal then the general theory 
of multipole moments to track the contribution of dipole moment to the current multipole moments.
If the expression for the Keplerian frequency [Eq.~(21)] and for the angular momentum [Eq.~(23)] in 
\cite{Sanabria10} are analyzed in detail, one finds that the signs of the contributions of the angular 
momentum and the magnetic dipole of the source coincide, this being said, one could argue that
the contribution of the dipole moment is related to an enhancement of the current multipole moment 
of the source.
This remark is confirmed by the general multipole expansion discussed by \cite{2004CQGra..21.5727S}. 
In particular, the influence of the dipole moment in the higher current-multipole moments is clear from 
Eq.~(23)--(25) of this reference.

Hence, the role of the dipole moment of the source in the dynamics of neutral test particles is 
qualitatively analogous to the role of the angular momentum, albeit it induces corrections of higher 
orders in multipole expansion of the effective potential.
This is a subject that deserves a detailed discussion and will be explained somewhere else.

%

%

\section{ Concluding Remarks}
\label{sec:con} 
%
The influence of weak electromagnetic fields on the dynamics of test particles around astrophysical 
objects has mainly been studied for the case of orbiting charged particles.
Based on a analytic solution of the Einstein-Maxwell field equations, here it is shown that a strong 
magnetic field, via the energy mass relation, modifies the dynamics of \emph{neutral} test particles. 
In particular, it is shown that an intense magnetic field induces corrections in the Keplerian, precession 
and oscillation frequencies of the radial and vertical motions of the test particles as well as in the 
dragging of inertial frames.

In particular, it was shown that if the angular momentum and the dipole moment are parallel (note that 
$j$ and $\mu$ have the same sign), then the ISCO radius of co-rotating orbits decreases for increasing
dipole moment, this leads to an increase of the Keplerian, and precession and oscillation frequencies 
of the radial and vertical motions frequency.
The angular momentum and the energy of the ISCO decrease for increasing dipole moment as a
consequence of the dragging of inertial frames. 

The kind of geodetical analysis performed here is widely used for instance, in the original RPM model 
\citep{Stella99,1999ApJ...524L..63S}, and int its subsequent reformulations, of the HF QPOs 
observed in LMXBs.
However, \cite{2011ApJ...726...74L} and \cite{0004-637X-760-2-138} have concluded that although these 
models of HF QPOs, which neglect  the influence of strong magnetic fields, are qualitatively satisfactory, 
they do not provide satisfactory fits to the observational data.
Hence, in order to improve (i) the level of physical description of these models and (ii) the fit to 
observational data, a more detailed analysis on the role of the extremely strong magnetic field 
in the structure of the spacetime is necessary and will be performed in the future.

\section{Acknowledgments}  
This work was partially supported by Fundaci\'on para la Promoci\'on de la Investigaci\'on y la 
Tecnolog\'ia del Banco la Rep\'ublica grant number 2879. 
C.A.V.-T. is supported by Vicerrector\'ia de Investigaciones (UniValle) grant number 7859.
L.A.P. acknowledges the financial support by the \textit{Comit\'e para el Desarrollo de la Investigaci\'on} 
--CODI-- of Universidad de Antioquia.

\bibliographystyle{AD2} 
\bibliography{geodesics}

\end{document}